\documentclass[journal=jacsat,manuscript=article]{achemso}

\usepackage[T1]{fontenc} 
\usepackage{float}       
\usepackage{helvet}      
\usepackage{mathptmx}    
\usepackage{setspace}    

\usepackage{bm}
\usepackage{graphicx}
\usepackage{amssymb}
\usepackage{amsmath}
\author{Adri\'{a}n Francisco-L\'{o}pez}
\affiliation[ICMAB-CSIC]
{Institut de Ci\`encia de Materials de Barcelona (ICMAB-CSIC), Campus UAB, 08193 Bellaterra, Spain}
\author{Bethan Charles}
\author{Oliver J.~Weber}
\affiliation[University of Bath]
{Dept. of Chemistry \& Centre for Sustainable Chemical Technologies, University of Bath, Claverton Down, Bath BA2 7AY, UK}
\author{M.~Isabel Alonso}
\author{Miquel Garriga}
\author{Mariano Campoy-Quiles}
\affiliation[ICMAB]
{Institut de Ci\`encia de Materials de Barcelona (ICMAB-CSIC), Campus UAB, 08193 Bellaterra, Spain}
\author{Mark T.~Weller}
\affiliation[University of Bath]
{Dept. of Chemistry \& Centre for Sustainable Chemical Technologies, University of Bath, Claverton Down, Bath BA2 7AY, UK}
\author{Alejandro R.~Go\~ni}
\email{goni@icmab.es}
\affiliation[ICMAB]
{Institut de Ci\`encia de Materials de Barcelona (ICMAB-CSIC), Campus UAB, 08193 Bellaterra, Spain}
\alsoaffiliation{ICREA, Passeig Llu\'is Companys 23, 08010 Barcelona, Spain}

\title{On the role of the electron-phonon interaction in the temperature dependence of the gap of lead halide perovskites}

\begin{document}

\begin{abstract}

Lead halide perovskites are causing a change of paradigm in photovoltaics. Among other peculiarities, these perovskites exhibit an atypical temperature dependence of the fundamental optical gap: It decreases in energy with decreasing temperature. So far reports ascribe such a behavior to a particularly strong electron-phonon renormalization of the band gap, neglecting completely contributions from thermal expansion effects. However, high pressure experiments performed, for instance, on the archetypal perovskite MAPbI$_3$, where MA stands for methylammonium, yield a negative pressure coefficient for the gap of the tetragonal room-temperature phase, which speaks against the assumption of a negligible gap shift due to thermal expansion. On the basis of the high pressure results, we show here that for MAPbI$_3$ the temperature-induced gap renormalization due to electron-phonon interaction can only account for about 40\% of the total energy shift, thus implying thermal expansion to be the dominant term. Furthermore, this result possesses general validity, holding also for the tetragonal or cubic phase, stable at ambient conditions, of other halide perovskite counterparts.

\end{abstract}

\noindent


Hybrid lead halide perovskites of the type APbX$_3$ with organic A-site cation and halide substitution on the X site are being the focus of attention of photovoltaic research and to a lesser extent as light emitters. These materials which can be cost-effectively deposited from solution, experienced a rash improvement in solar-energy conversion efficiency, recently reaching a remarkable value of 23.7\% \cite{nrelx19a}. For a solar cell as well as for a light emitting device the band gap and its temperature dependence are fundamental properties of the active material. A peculiarity of the tetragonal and cubic phases of hybrid lead halide perovskites, which are stable at ambient conditions, is that they exhibit an {\it atypical} dependence on temperature of the fundamental direct gap: it decreases in energy with decreasing temperature. This temperature dependence of the gap is almost ubiquitous in halide perovskites, for example, in MAPbI$_3$ \cite{foley15a,milot15a,darxx16a,wrigh16a,kimxx17a}, MAPbBr$_3$ \cite{darxx16a,tilch16a,wrigh16a}, MAPbCl$_3$ \cite{wuxxx14a}, FAPbI$_3$ \cite{wrigh16a}, FAPbBr$_3$ \cite{darxx16a,wrigh16a}, FA$_x$MA$_{1-x}$PbI$_3$ \cite{franc19a}, CsPbI$_3$ \cite{saran17a} and CsPbBr$_3$ \cite{shind17a,saran17a}. By atypical it is meant opposite to the temperature behavior of the gaps exhibited by most of covalent bonded semiconductors, for which the gap increases with decreasing temperature (see, for instance, Refs. \cite{laute85a,gopal87a,laute87a,cardo89a,cardo05a} and references therein).

The theoretical framework for describing the variation of band gaps with temperature was set by V. Heine and P.B. Allen \cite{allen76a} and further developed by M. Cardona and coworkers using the empirical pseudopotential method at an early stage \cite{allen81a,laute85a,gopal87a,cardo89a} and ab-initio techniques afterwards \cite{cardo05a,goebe98a,bhosa12a}. The changes of the semiconductor band structure with temperature arise essentially from the effect of thermal expansion (TE) due to the anharmonicity of the crystal potential and from electron-phonon interaction. The renormalization of the band energies due to electron-phonon coupling, when considered to second order in the atomic displacements, consists of two terms: The Debye-Waller (DW) and the self-energy (SE) corrections. Usually, for the direct gaps of most semiconductors both terms cause a reduction with increasing temperature. The decrease caused by thermal expansion is mainly due to a positive hydrostatic pressure coefficient of the gaps, whereas the gap reduction arising from electron-phonon interaction is proportional to the Bose-Einstein phonon occupation number. There are, however, exceptions to the mentioned rule. The first reported material which appeared to show an abnormal temperature dependence of the fundamental gap is CuCl, which exhibits a slight sublinear increase of the gap with increasing temperature \cite{goebe98a}. Other cuprous halides \cite{serra02a} and several silver chalcopyrites like AgGaS$_2$ and AgGaSe$_2$ also display abnormal behavior but at low temperatures \cite{bhosa12a}. A common characteristic of copper halides and silver chalcopyrites is that the pressure coefficient of the gap is very small, such that the thermal expansion contribution to the gap renormalization becomes almost irrelevant. As a consequence, in these cases, the temperature dependence of the gap is mainly determined by electron-phonon interaction. Probably mislead by these early results, this point of view was generally adopted to interpret the atypical temperature dependent renormalization of the fundamental gap of lead halide perovskites exclusively as due to a particularly strong electron-phonon coupling \cite{tilch16a,saran17a,saidi18a}.

In this Letter, we demonstrate that the role of the electron-phonon interaction in the temperature-induced renormalization of the gap of lead halide perovskites has been widely overestimated so far. Using available data for the hydrostatic pressure coefficient of the gap, we show for the archetypal perovskite compound MAPbI$_3$ that the thermal expansion contribution is the leading term in the temperature dependence of the direct gap, whereas electron-phonon interaction effects accounts only for ca. 40\% of the total energy shift. Moreover, we provide arguments in favor of an interpretation of the sign and magnitude of the electron-phonon renormalization terms, which holds also for other halide perovskites. Based on the results of previous empirical pseudopotential calculations, we infer that the electron-phonon coupling affects electronic states in different ways, depending on their bonding/antibonding and atomic orbital character.


The MAPbI$_3$ samples used for the high pressure experiments are high quality single crystals grown by the space-confined on-substrate fabrication method, as reported elsewhere \cite{franc18a}, having a suitable final thickness of ca. 30 $\mu$m. Otherwise, for the experiments as a function of temperature we used large and thick single crystals of about $2\times2\times1$ mm$^3$ synthesized from aqueous solution by similar procedure \cite{pogli87a}. The high-pressure photoluminescence (PL) measurements were performed at room temperature employing a gasketed diamond anvil cell (DAC) with anhydrous propanol as pressure transmitting medium  \cite{franc18a}, whereas the temperature dependent PL measurements were carried out in vacuum using a gas-flow cryostat  \cite{franc19a}. The PL spectra were excited with the 633 nm line of a He-Ne laser using a very low incident light power density below 15 W/cm$^2$ to avoid any photo-degradation of the samples. Spectra were collected using a 20$\times$ long working distance objective with NA=0.35 and dispersed with a high-resolution LabRam HR800 grating spectrometer equipped with a charge-coupled device detector. PL spectra were corrected for the spectral response of the spectrometer by normalizing each spectrum using the detector and the 600-grooves/mm grating characteristics.


\begin{figure}[htbp]
\includegraphics[width=6cm]{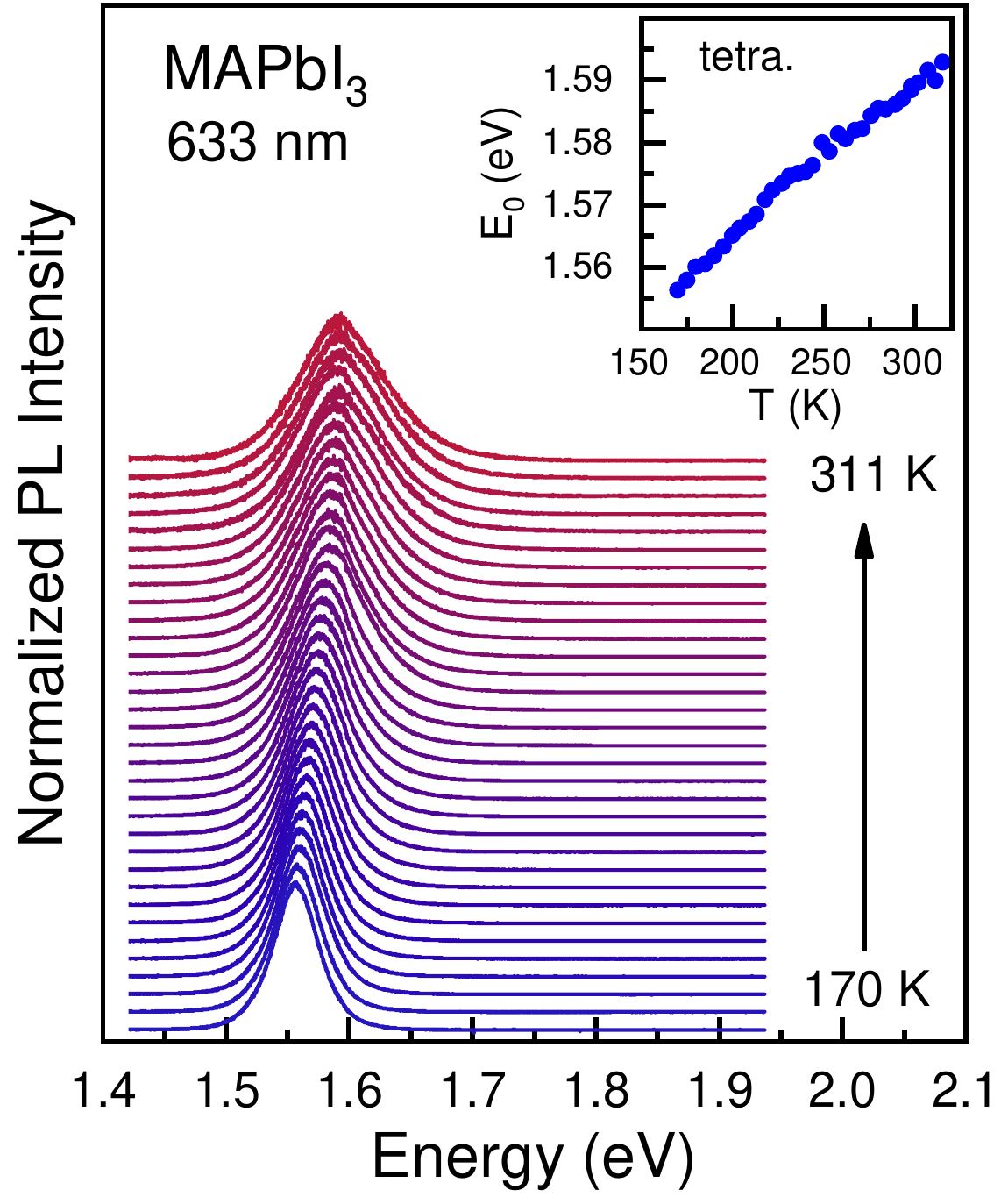}
\caption{
\label{PL-spectra-T}
PL spectra of MAPbI$_3$ measured at different temperatures in the range of stability of the tetragonal phase (ca. 170 to 311 K) using the red line (633 nm) for excitation. The spectra were normalized to their maximum intensity and plotted with a vertical shift for increasing temperature. The inset shows the temperature dependence of the maximum peak position $E_0$ of the spectra displayed in the body of the figure.
}
\end{figure}

Figure \ref{PL-spectra-T} shows the evolution of the PL spectra with temperature in the stability range of the tetragonal phase of MAPbI$_3$. All spectra were normalized to its absolute maximum intensity and vertically offset for clarity. The main PL peak exhibits a gradual redshift and sharpening with decreasing temperature. To analyze the PL spectra of MAPbI$_3$ we used a Gaussian-Lorentzian cross-product function for describing the main peak which is ascribed to free-exciton recombination \cite{franc18a}. The values of the fitting parameter corresponding to the energy $E_0$ of the exciton peak maximum are plotted as a function of temperature in the inset to Fig. \ref{PL-spectra-T}. Although we cannot tell the absolute values of bandgap and/or exciton binding energy from the lineshape fits, the shift of the PL peak energy $E_0$ with temperature (or pressure) is to a large extent dictated by the shift of the gap which for MAPbI$_3$ exhibits a fairly linear decrease with decreasing temperature. 

\begin{figure}[htbp]
\includegraphics[width=7cm]{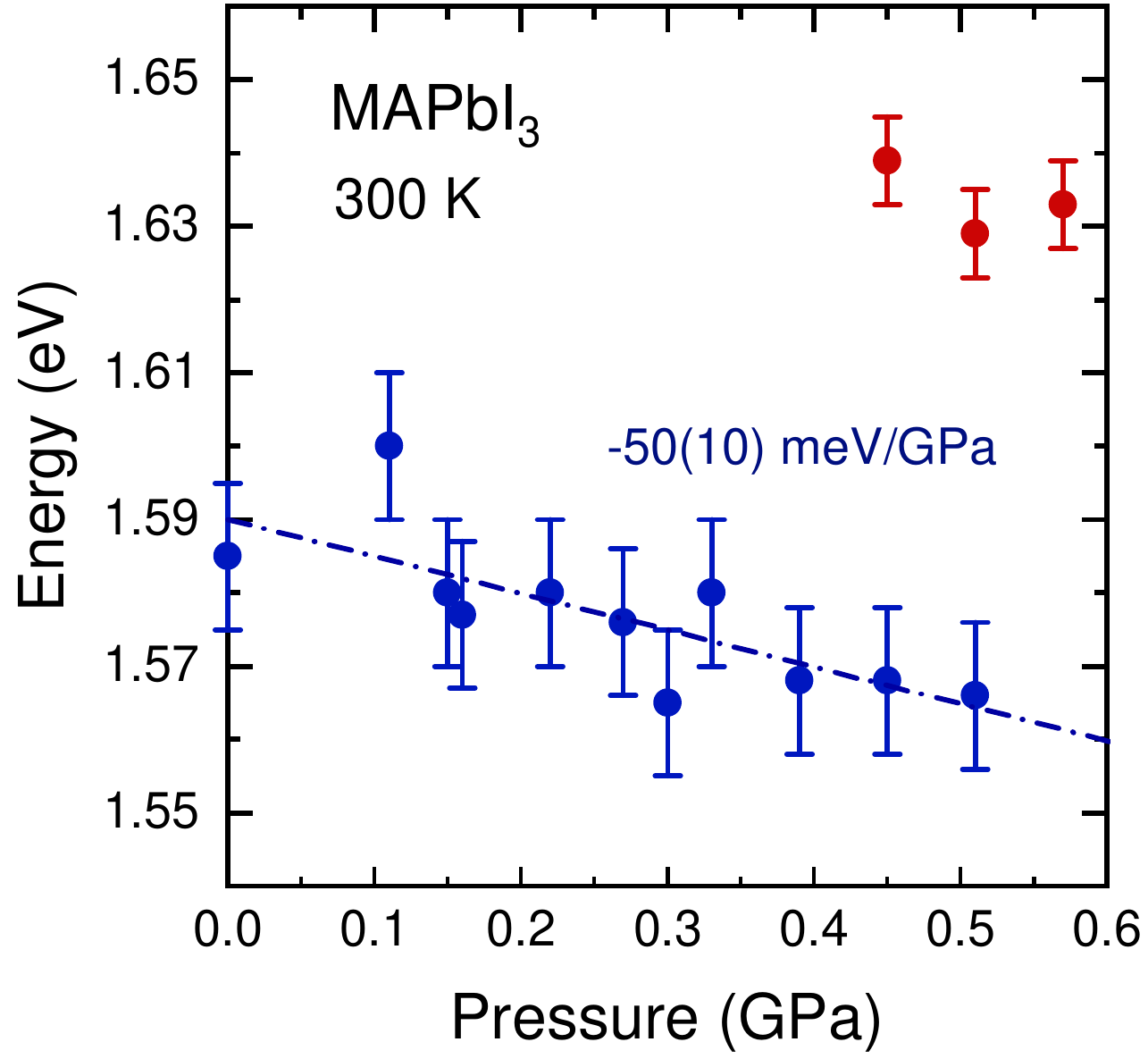}
\caption{
\label{Eg-P}
The energy $E_0$ of the PL peak maximum (blue symbols) plotted as a function of pressure in the stability range of the tetragonal phase of MAPbI$_3$. The dot-dashed line represents a fit to the data points using a linear function (its slope is indicated). The red symbols correspond to PL emission from regions of the sample which had already underwent the phase transition into the high pressure cubic phase (see Ref. \cite{franc18a} for details).
}
\end{figure}

As mentioned before, the derivative of the gap over temperature contains two terms; one accounts for thermal expansion effects (TE) and the other corresponds to the renormalization directly caused by electron-phonon interaction (EP), which includes the Debye-Waller and self-energy corrections \cite{laute85a,gopal87a,goebe98a}:
\begin{equation}
\frac{dE_g}{dT}=\left[\frac{\partial E_g}{\partial T}\right]_{TE}+\left[\frac{\partial E_g}{\partial T}\right]_{EP}.
\label{dEg/dT}
\end{equation}
\noindent The effect on the gap due to the contraction of the lattice with decreasing temperature is intimately related to the response of the electronic band structure upon application of external hydrostatic pressure. It thus holds \cite{laute85a,gopal87a,cardo05a}
\begin{equation}
\left[\frac{\partial E_g}{\partial T}\right]_{TE}=-\alpha_V\cdot B_0\cdot\frac{dE_g}{dP},
\label{TE}
\end{equation}
\noindent where $-\alpha_V$ is the volumetric expansion coefficient, $B_0$ is the bulk modulus, i.e. the inverse of the compressibility, and $\frac{dE_g}{dP}$ is the pressure coefficient of the gap, which can be determined from high pressure experiments. The last two factors depend only weakly on temperature. As shown in the Supplementary Information, the strongest temperature variation comes from $\alpha_V$ (also including zero-point vibrations \cite{cardo05a}). At room temperature $\alpha_V$ is positive and the sign of Eq. (\ref{TE}) for the thermal expansion contribution is determined by the sign of the pressure coefficient. For most semiconductor direct gaps $\frac{dE_g}{dP}$ is positive as well. Hence, thermal expansion causes a gap reduction. The lead halide perovskites are an exception. Figure \ref{Eg-P} shows the variation with pressure of the energy position of the PL peak measured in MAPbI$_3$ single crystals with the DAC in the short stability range of the tetragonal phase. A linear fit to the data points (dot-dashed line) yields an unusually large but negative pressure coefficient, as indicated (the negative sign of $\frac{dE_g}{dP}$ arises from the inverted atomic orbital character of the states at the top and bottom of the valence and conduction band of MAPbI$_3$, respectively, as explained elsewhere \cite{franc18a}). This implies that for MAPbI$_3$ the expansion of the lattice with increasing temperature leads to a gradual opening of the gap, as displayed in the inset to Fig. \ref{PL-spectra-T}. The magnitude of this effect will be discussed later together with the gap renormalization due to electron-phonon interaction.

Gopalan {\it et al.}\cite{gopal87a} derived an expression for the shift and broadening induced by temperature, through electron-phonon interaction, of any electronic state $E_{n\mathbf{k}}$ with band index $n$ and wavevector $\mathbf{k}$, for which all phonon modes of the branch $j$, wavevector $\mathbf{q}$ and frequency $\omega_{j\mathbf{q}}$ contribute:
\begin{equation}
\Delta E_{n\mathbf{k}}(T)= \sum_{j\mathbf{q}}\frac{\partial E_{n\mathbf{k}}}{\partial n_{j\mathbf{q}}}\left(n_{j\mathbf{q}}(T)+\frac{1}{2}\right),
\label{renorm1}
\end{equation}
\noindent where $n_{j\mathbf{q}}=\left(e^{\beta\hslash\omega_{j\mathbf{q}}}-1\right)^{-1}$ is the Bose-Einstein phonon occupation factor with $\beta=\frac{1}{k_BT}$. The real part of the complex interaction coefficients $\frac{\partial E_{n\mathbf{k}}}{\partial n_{j\mathbf{q}}}$ contribute to the energy shift of the bands and contain both the DW and SE parts, whereas the imaginary part leads to a lifetime broadening of the electronic states. Obviously, the sign of the renormalization of a gap is thus determined by the difference in magnitude and sign of the respective energy shift of valence and conduction band.

By invoking energy conservation, the summation in Eq. (\ref{renorm1}) transforms into an integral over the phonon frequencies \cite{gopal87a}:
\begin{equation}
\begin{split}
\Delta E_{n\mathbf{k}}(T) & = \int_0^\infty d\omega\cdot g^2F(n,\mathbf{k},\omega)\cdot\left(n_{j\mathbf{q}}(T)+\frac{1}{2}\right) \\
g^2F(n,\mathbf{k},\omega) & = \sum_{j\mathbf{q}}\frac{\partial E_{n\mathbf{k}}}{\partial n_{j\mathbf{q}}}\delta(\omega-\omega_{j\mathbf{q}}).
\end{split}
\label{renorm2}
\end{equation}
\noindent The function $g^2F(n,\mathbf{k},\omega)$ is the so-called electron-phonon spectral function and is essentially the phonon density of states (DOS) appropriately weighted by electron-phonon matrix elements. As such, the spectral function is temperature independent, which means that the temperature dependence of the electron-phonon contribution to the gap shift arises solely from the Bose-Einstein occupation factor $n_{j\mathbf{q}}(T)$.

This is an important result because Eqs. (\ref{renorm1}) and (\ref{renorm2}) clearly indicate that the main contributions to the electron-phonon renormalization of the gap arise from peaks in the phonon density of states. In fact, this is at the origin of the Einstein-oscillator model introduced by Cardona and coworkers \cite{goebe98a,serra02a,bhosa12a}, which approximate the $\frac{\partial E_{n\mathbf{k}}}{\partial n_{j\mathbf{q}}}$ coefficients by effective electron-phonon interaction parameters $A_i$ for phonons with average frequency $\omega_i$, inferred from the peaks in the phonon DOS. The EP correction to the gap then reads
\begin{equation}
\left[\Delta E_g(T)\right]_{EP}= \sum_i A_i\cdot\left(n_B(\omega_i,T)+\frac{1}{2}\right),
\label{Einstein}
\end{equation}
\noindent where $n_B$ again stands for the Bose-Einstein factor. In special cases of materials with two atoms per unit cell with markedly different masses like the cuprous halides \cite{goebe98a,serra02a}, it is allowed to use a two-oscillator model with a modified effective EP coefficient which explicitly accounts for its dependence on the average phonon frequency and atomic specie mass ($A_i\rightarrow\frac{A'_i}{\omega_i\cdot M_i}$). This is justified because the phonon DOS exhibits two peaks, one at the average frequency of the acoustic phonon branches at the Brillouin zone edges, corresponding to vibrations of the heavier mass specie, and another oscillator accounting for the optical phonon contribution, corresponding to vibrations of the lighter mass atomic specie. For exactly the same reason each of the two oscillators is identified with the contribution from acoustic and optical phonons to the electron-phonon renormalization. We point out that this model cannot be simply transferred to the case of the halide perovskites \cite{tilch16a,saran17a}, because perovskites have many atoms per unit cell and the vibrations cannot be classified as lead only or halide only. In fact, an inspection of the phonon DOS for the three methylammonium lead halide compounds \cite{leguy16a} indicates that the DOS exhibits up to four well-defined peaks or bands, not two, with partial intermixing of optical and acoustical branches.

\begin{figure}
\includegraphics[width=7cm]{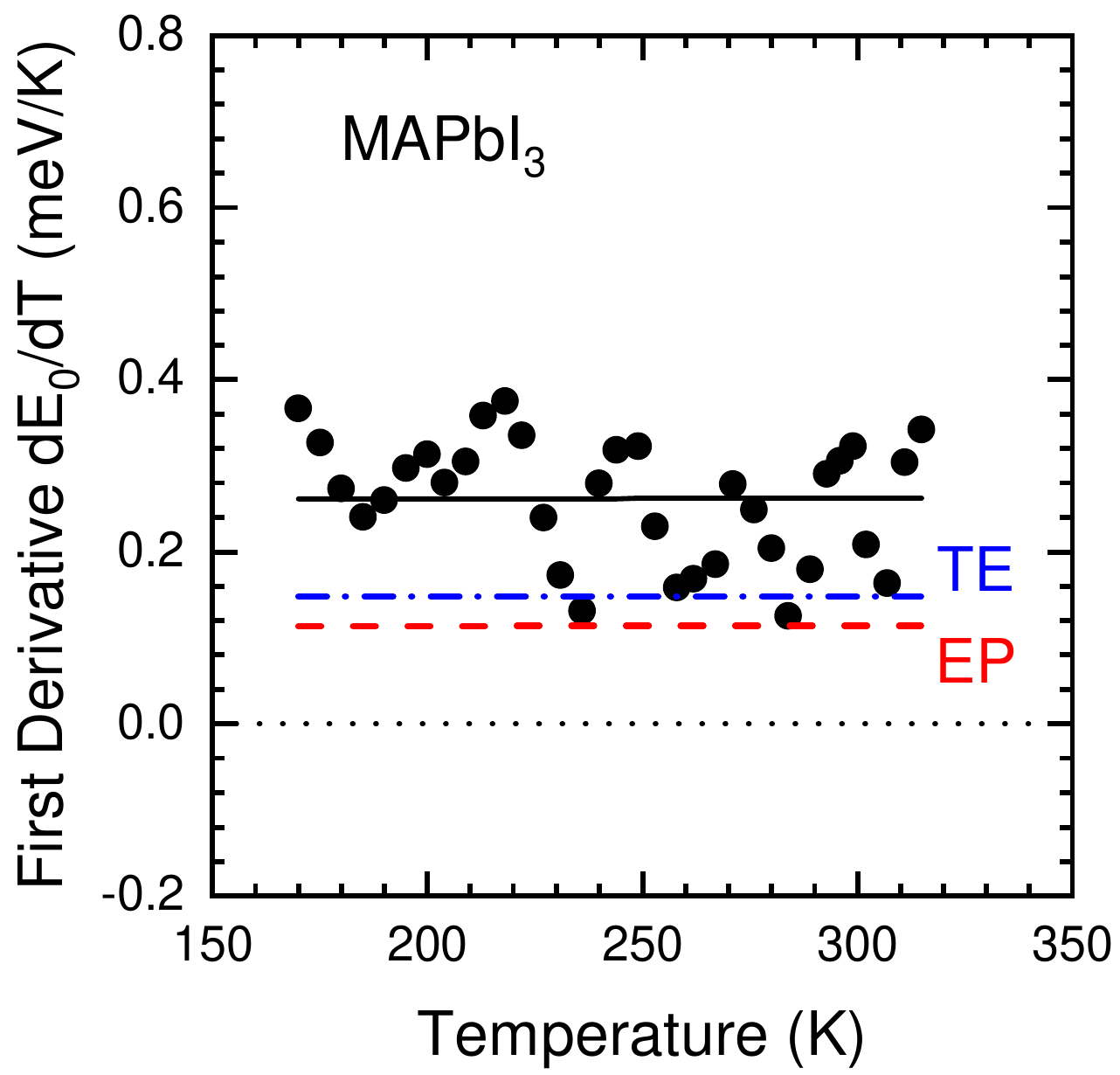}
\caption{
\label{dE-dT}
The first derivative of the exciton peak energy $E_0$ with respect to temperature (closed black symbols), numerically calculated from the data of the inset to Fig. \ref{PL-spectra-T}. The solid black line represents a fit to the data points, corresponding to the sum of the contribution of thermal expansion (blue dot-dashed curve) and electron-phonon interaction (red dashed curve). See text for details.
}
\end{figure}

In this respect, we show here that the electron-phonon renormalization can be well accounted for using a single Einstein oscillator with a {\it positive} effective coefficient $A_{eff.}$, so as to reproduce the linear decrease of the gap with decreasing temperature of MAPbI$_3$. Since we do not know the absolute magnitude of the gap renormalization at room temperature (and any other temperature), we decided to circumvent this handicap by evaluating the derivative with respect to temperature of the gap renormalization instead. The closed black symbols in Fig. \ref{dE-dT} represent the first derivative over temperature of the bandgap, numerically calculated from the data points of $E_0$ versus T shown in the inset to Fig. \ref{PL-spectra-T}. Despite the dispersion of the values, in the temperature range of stability of the tetragonal phase of MAPbI$_3$, the first derivative of the gap over temperature is, as expected, essentially constant. The contribution from thermal expansion can be directly obtained from Eq. (\ref{TE}), using the values of $\alpha_V=1.57\times10^{-4}$ K$^{-1}$ \cite{jacob15a} and $B_0=18.8$ GPa \cite{szafr16a} from the literature and the linear pressure coefficient determined by us from high pressure experiments, indicated in Fig. \ref{Eg-P}. The result is the constant contribution represented by the blue dot-dashed line in Fig. \ref{dE-dT}. In order to calculate the contribution from electron-phonon interaction, we have derived Eq. (\ref{Einstein}) with respect to temperature, considering a single oscillator with coupling constant $A_{eff.}$ and oscillator frequency $\omega_{eff.}$:
\begin{equation}
\left[\frac{dE_g}{dT}\right]_{EP}= \frac{A_{eff.}}{4T}\cdot\frac{\hslash\omega_{eff.}}{k_BT}\cdot\frac{1}{sinh^2\left(\frac{\hslash\omega_{eff.}}{2k_BT}\right)}.
\label{EP}
\end{equation}
\noindent This function together with the constant contribution from TE has been fitted to the data points of Fig. \ref{dE-dT} using only the electron-phonon coupling constant and the average phonon frequency as adjustable parameters. The resulting values for these parameters are: $A_{eff.}=8.0(8)$ meV and $\hslash\omega_{eff.}=5.8(6)$ meV, where the numbers in parentheses are the error bars (uncertainty of the last digits). The solid black curve and the dashed red curve in Fig. \ref{dE-dT} represent the total rate of gap renormalization per Kelvin and the EP contribution to it, respectively. At 300 K we obtain a total renormalization rate of 0.26(5) meV/K, which splits into 0.15(5) meV/K from thermal expansion and 0.11(5) meV/K from electron-phonon interaction. This is a clear indication that thermal expansion effects (58\%), rather than being negligible \cite{tilch16a,saran17a,saidi18a}, are even more important than electron-phonon coupling effects (42\%). We show below that this situation also holds for most, if not all, lead halide perovskites investigated so far.

By accounting for the thermal expansion effects, we obtain within the Einstein-oscillator model a value of ca. 6 meV for effective phonon frequency involved in the gap renormalization, which is in excellent agreement with the frequency of 1 THz (approx. 4 meV) of the specific phonon mode which strongly couples to the gap in THz transient transmission experiments \cite{kimxx17a} and with the frequency of 4.2(8) meV of the phonons leading to exciton broadening \cite{diabx16a} both in MAPbI$_3$, although the latter holds for the orthorhombic phase. In contrast, if the two-oscillator model is applied without thermal expansion consideration, unrealistically large average phonon frequencies for the optical branches of approx. 16 meV \cite{tilch16a} or even 40 to 50 meV \cite{saran17a} are found for the best fit to the gap-versus-temperature data. We recall that the maximum cutoff frequency of the phonon modes of the inorganic cages of lead halide perovskites is 200 cm$^{-1}$ (ca. 25 meV) \cite{brivi15a} but the most intense Raman modes have frequencies below 100 cm$^{-1}$ (ca. 12.5 meV) \cite{leguy16a}.

\begin{figure}
\includegraphics[width=7cm]{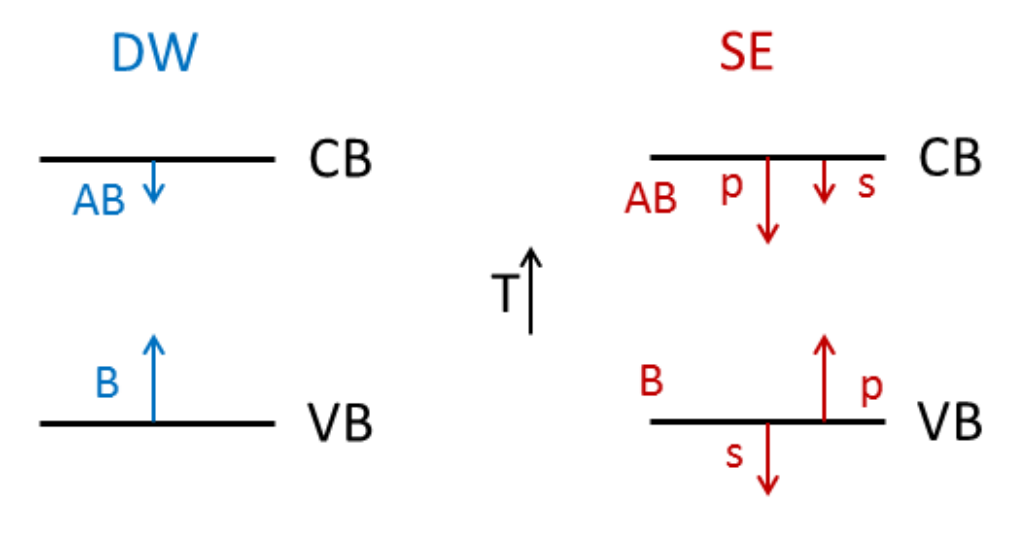}
\caption{
\label{EPP}
Sketch of the sign and magnitude (arrows) of the Debye-Waller (DW) and self-energy (SE) contributions to the electron-phonon renormalization of valence and conduction band states of $sp^3$ bonded semiconductors, depending on the bonding (B) or antibonding (AB) and atomic orbital character ($s$ or $p$) of the involved states for the case of increasing temperature (T).
}
\end{figure}

The calculation by ab-initio methods of the magnitude of the electron-phonon interaction in materials like the hybrid halide perovskites, characterized by soft and strongly anharmonic vibrational modes, constitutes a real challenge \cite{saidi16a,wrigh16a,whall16a,whall17a}. Despite their relevance, we prefer to make use of the semiempirical pseudopotential calculations that provided a detailed, though qualitative, picture of the electron-phonon interaction in covalent, $sp^3$-bonded semiconductors \cite{allen81a,laute85a,gopal87a,cardo89a} (for further details see the Supplementary Information). The Debye-Waller correction represents two-phonon processes for which the lattice vibrations perturb the electronic band structure in the same way as in the case of x-ray diffraction patterns, i.e., by smearing the pseudopotential {\it structure factor}. In this sense, the DW correction mainly depends on the spatial distribution of the electronic charge around the lattice atoms. As illustrated in the sketch of Fig. \ref{EPP}, valence band states with mostly bonding character increase in energy with increasing temperature, whereas the antibonding conduction band states slightly decrease in energy. Hence, the DW term always cause a gap reduction with increasing temperature for sp$^3$-bonded semiconductors. In contrast, the sign and magnitude of the self-energy term would depend on how the pseudopotential {\it form factors} react to the different phonon eigenvectors corresponding to the modes leading to the peaks in the phonon DOS. It turns out that the acoustic-phonon SE contribution almost cancels out the DW correction \cite{gopal87a}, which implies that for sp$^3$ bonding only optical phonon contribute to the gap renormalization. The remaining SE interaction matrix elements are large and positive for $p$-like and moderate but negative for $s$-like bonding valence-band states, whereas for antibonding conduction-band states the SE interaction parameters are negative for $s$ and $p$-like states but larger for the latter (see Fig. \ref{EPP}). As a consequence a single Einstein oscillator with a negative coupling constant $A_{eff.}$ provides a good description of the gap reduction induced by an increase in temperature in conventional semiconductors.

A common feature of the materials showing an abnormal temperature dependence of their gap, like certain cuprous halide \cite{goebe98a,serra02a} and Ag chalcopyrite \cite{bhosa12a} compounds, is the presence of $d$-states in the valence band. The Cu $3d$ and Ag $4d$-like valence electrons hybridize with the usual $p$-like counterparts, reverting the sign of the electron-phonon interaction and so leading to a strong downshift of the valence band with increasing temperature. The $d$-states hybridization also has important consequences for the pressure coefficient of the gap \cite{weixx98a}, which is largely reduced, because the AB $d$-states push up the top of the valence band with pressure at a similar pace than the upshift of the bottom of the conduction band. On the contrary, the unusual temperature-induced gap renormalization in lead halide perovskites possesses a different origin, since there is no hybridization with $d$-states whatsoever. It is the huge spin-orbit interaction which causes a so-called band inversion. Relativistic band-structure calculations \cite{giorg13a,frost14a,evenx15a} for a pseudo-cubic phase of MAPbI$_3$ predict that for the direct gap at the R-point of the Brillouin zone the top of the valence band is predominantly composed by Pb $6s$ orbitals slightly hybridized with I $5p$ orbitals, whereas the bottom of the conduction band is formed by the split-off Pb $6p$-orbitals. Hence, one expects the EP interaction to lead to a gap increase with increasing temperature (see Suplementary Information for details). This is supported by ultraviolet photoemission spectroscopy combined with optical gap measurements, which for MAPbI$_3$ show a stronger lowering of the valence band maximum with respect to the conduction band minimum, as temperature is raised \cite{foley15a}. Furthermore, such band inversion also explains its negative pressure coefficient \cite{franc18a}. The latter determines the sign of the thermal expansion contribution, which thus adds up to the effects of electron-phonon interaction as far as the gap renormalization is concerned. Since all arguments presented here are valid for all halide perovskites crystallizing in the tetragonal and/or cubic phases, the same behavior of the gap with temperature, which is ubiquitous in this material system, has similar explanation.


In conclusion, we have shown that the importance of the electron-phonon interaction in the {\it abnormal} temperature dependence of the fundamental gap of the tetragonal or cubic phases of lead halide perovskites has been widely overestimated in previous work. This was the result of totally neglecting the effects of thermal expansion. As a consequence, disproportionately large electron-phonon coupling constants and average phonon frequencies were needed to explain the variation of the gap with temperature in halide perovskites. Using MAPbI$_3$ as representative example, we showed that the thermal expansion effects can be readily quantified from the measured (also unusual) negative pressure coefficient of the gap. Our findings, which have general validity for lead halide perovskites, clearly indicate that thermal expansion has to be treated on equal footing with the electron-phonon interaction for the correct interpretation of temperature effects on the electronic structure of halide perovskites. Given the relevance of the electron-phonon interaction for a variety of physical phenomena apart from the temperature dependence of the gap (charge transport, exciton lifetimes, non-radiative relaxation processes, thermoelectric properties, etc.), its correct assessment is fundamental for further scientific and/or technological developments with lead halide perovskites.

\section{Supporting Information} Contains details of the theoretical discussion, based on the empirical pseudopotential method, of the effects of thermal expansion and electron-phonon interaction on the renormalization of gaps with temperature for conventional semiconductors.

\section{Acknowledgements}

We gratefully acknowledge fruitful discussions with H. M\'{\i}guez, M. Calvo and A. Rubino from the Institute of Materials Science of Seville, Spain.
The Spanish Ministerio de Ciencia, Innovaci\'{o}n y Universidades is gratefully acknowledged for its support through Grant No. SEV-2015-0496 in the framework of the Spanish Severo Ochoa Centre of Excellence program and through Grant MAT2015-70850-P (HIBRI2). AFL acknowledges a FPI fellowship from the Spanish Ministerio co-financed by the European Social Fund and the PhD programme in Materials Science from Universitat Aut\`{o}noma de Barcelona in which he is enrolled. BC and OJW thank the EPSRC for PhD studentship funding via the CSCT CDT (EP/G03768X/1, EP/L016354/1). Financial support from is also acknowledged from the European Research Council through project ERC CoG648901.


\end{document}


\section{Relation between thermal expansion (TE) and Einstein oscillator model}

From thermodynamical arguments \cite{ashcr76a,cardo05a} it is possible to relate the volumetric thermal expansion with the Gr\"{u}neisen parameter $\gamma_{j\mathbf{q}}=-\frac{\partial ln\omega_{j\mathbf{q}}}{\partial ln V}$ and Bose-Einstein occupation number $n_{j\mathbf{q}}(T)$ of all phonon modes as
\begin{equation}
\begin{split}
\frac{\Delta V(T)}{V_0} & = \alpha_V\cdot\Delta T \\
& = \frac{1}{B_0\cdot V_0}\cdot\sum_{j\mathbf{q}}\hslash\omega_{j\mathbf{q}}\gamma_{j\mathbf{q}}\left(n_{j\mathbf{q}}(T)+\frac{1}{2}\right),
\end{split}
\label{Gruneisen}
\end{equation}
\noindent where the summation runs over all phonon modes of the branch $j$, wavevector $\mathbf{q}$ and frequency $\omega_{j\mathbf{q}}$. Two important results might be surmised from Eq. (\ref{Gruneisen}). On the one hand, we note that the electron-phonon coupling is somehow indirectly contained in the thermal expansion (TE) contribution through the phonon Gr\"{u}neisen parameters. On the other hand, the temperature dependence of the TE gap renormalization is given by the Bose-Einstein occupation factor. This is the same dependence as for the Debye-Waller (DW) and self-energy (SE) contributions, directly accounting for electron-phonon interaction effects. Thus, if the TE contribution is neglected, i.e. not taken explicitly into consideration, an Einstein-oscillator model would nevertheless provide a good description of the temperature dependence of the bandgap, although at the expense of yielding unphysical values of the mean phonon frequency and/or electron-phonon coupling constant.

\section{The electron-phonon (EP) interaction and the empirical pseudopotential method}

For the discussion of the electron-phonon interaction it is strictly necessary to count with a satisfactory description of the electronic states as well as the spectrum of vibrations of the solid. In the former case, the empirical pseudopotential method has demonstrated to be a powerful tool for the calculation and understanding of the temperature dependence of direct band gaps in covalent semiconductors \cite{allen81a,laute85a,gopal87a,cardo89a}. Within the pseudopotential approximation, the electronic states correspond to the eigenvectors $\Psi_{n\mathbf{k}}$ with band index $n$ and wavevector $\mathbf{k}$, obtained from the following secular equations \cite{gopal87a}:
\begin{equation}
\begin{split}
& \sum_{\mathbf{G'}}\left[\left(\frac{\hslash^2}{2m}\cdot\left(\mathbf{k}+\mathbf{G}\right)^2-E_{n\mathbf{k}}\right)\cdot\delta_{\mathbf{G}\mathbf{G'}}+\sum_\kappa V_\kappa(\mathbf{G}-\mathbf{G'})\cdot S_\kappa(\mathbf{G}-\mathbf{G'})\right]\cdot C_{n\mathbf{k}}(\mathbf{G'})=0 \\
& \Psi_{n\mathbf{k}}\propto\sum_\mathbf{G}C_{n\mathbf{k}}(\mathbf{G})e^{i(\mathbf{k}+\mathbf{G})\cdot\mathbf{r}},
\end{split}
\label{EPP method}
\end{equation}
\noindent where $\mathbf{G},\mathbf{G'}$ are reciprocal-lattice vectors, $C_{n\mathbf{k}}(\mathbf{G})$ are the properly normalized Fourier coefficients of $\Psi_{n\mathbf{k}}$ and $V_\kappa(\mathbf{G})$ and $S_\kappa(\mathbf{G})$ are the pseudopotential form and structure factors, respectively, of the atom $\kappa$ in the unit cell. Rather than performing explicit matrix-element calculations, our aim is to attain a qualitative insight into the problem of the electron-phonon interaction that would allow us to draw general conclusions from a careful inspection of the empirical pseudopotential results available in the literature. In addition to the electronic states, a description of the vibrational properties of the solid are needed. Cardona and coworkers \cite{gopal87a,cardo89a} employed the rigid-ion model to calculate the phonon spectrum of semiconductors to obtain an expression for the electron-phonon (EP) renormalization of the electronic band-state energies as a function of temperature. All phonon modes of the branch $j$ and with wavevector $\mathbf{q}$ and frequency $\omega_{j\mathbf{q}}$ contribute to the renormalization of the electronic energies, which then reads as:
\begin{equation}
\Delta E_{n\mathbf{k}}(T)= \sum_{j\mathbf{q}\kappa}\frac{A(n,\mathbf{k},j,\mathbf{q},\kappa)}{\omega_{j\mathbf{q}}\cdot M_\kappa}\left(n_{j\mathbf{q}}(T)+\frac{1}{2}\right),
\label{EP-renorm}
\end{equation}
\noindent where $n_{j\mathbf{q}}=\left(e^{\beta\hslash\omega_{j\mathbf{q}}}-1\right)^{-1}$ is the Bose-Einstein phonon occupation factor with $\beta=\frac{1}{k_BT}$. In Eq. (\ref{EP-renorm}) we have explicitly written the dependence of the EP matrix element on the frequency $\omega_{j\mathbf{q}}$ of the phonon mode involved and the mass $M_\kappa$ of the vibrating atom (an average mass for modes with more than one atom vibrating). Equation (\ref{EP-renorm}) can be rewritten by transforming the summation into an integral over the phonon frequencies \cite{gopal87a}:
\begin{equation}
\Delta E_{n\mathbf{k}}(T) = \int_0^\infty d\omega\cdot g^2F(n,\mathbf{k},\omega)\cdot\left(n_{j\mathbf{q}}(T)+\frac{1}{2}\right). 
\label{EP-DOS}
\end{equation}
\noindent The function $g^2F(n,\mathbf{k},\omega)$ is the so-called electron-phonon spectral function and is essentially the phonon density of states (DOS) appropriately weighted by electron-phonon matrix elements. These matrix elements pick up contributions from two terms: The Debye-Waller (DW) and self-energy (SE) term (the corresponding electron-phonon Feynman diagrams, corresponding to second order perturbation in atomic displacements, are shown in Fig. 1 of Ref. \cite{gopal87a}, for instance).

\subsection{The Debye-Waller (DW) correction}

\begin{figure}
\includegraphics[width=7cm]{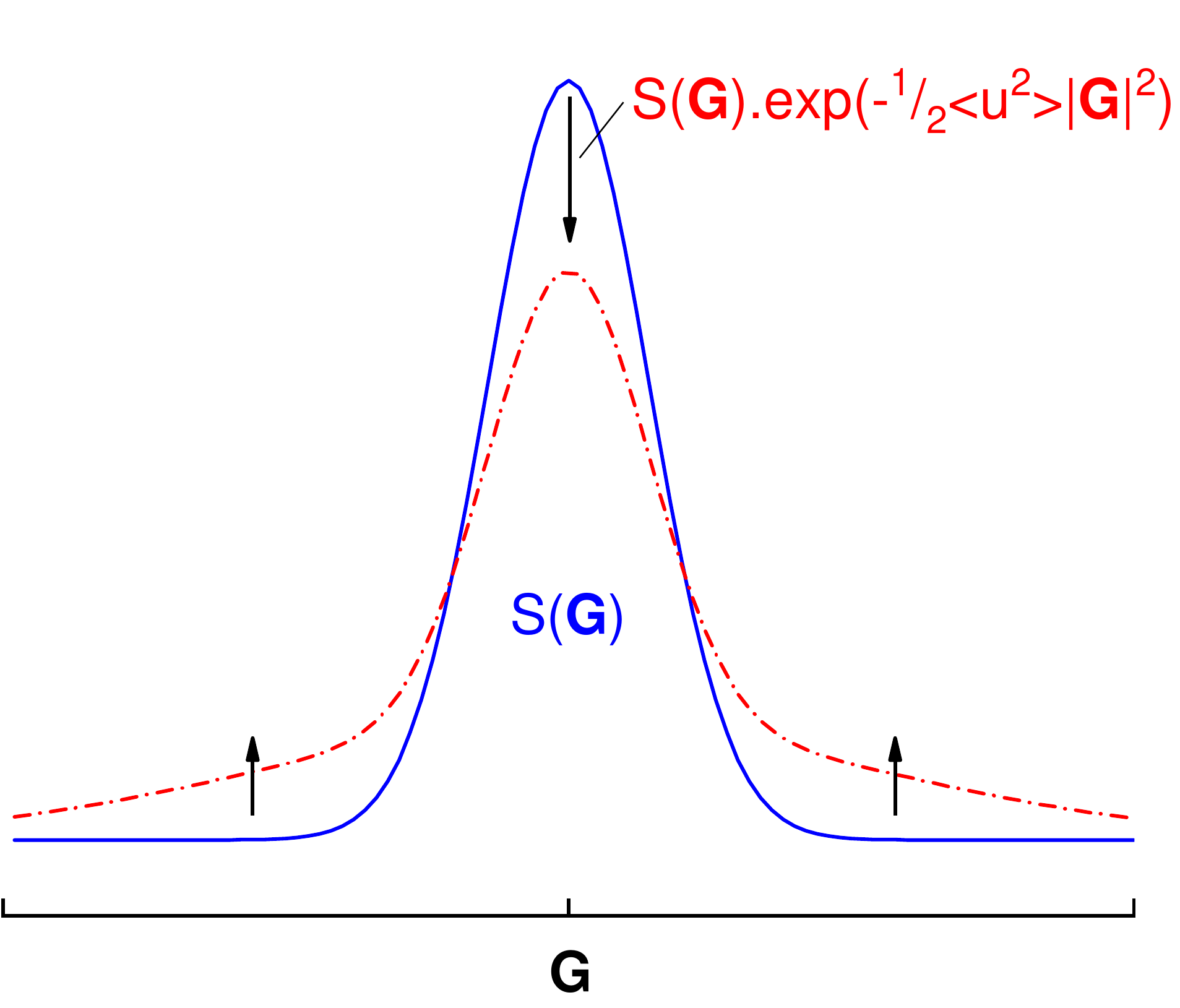}
\caption{
\label{DW-sketch}
Sketch showing the pseudopotential structure factor $S(\mathbf{G})$ (blue curve) in reciprocal space and the effect of the Debye-Waller factor upon it (red curve).
}
\end{figure}

The DW term accounts for the coupling between the lattice vibrations and the electronic band states through thermal fluctuations of the atoms around their equilibrium positions, which in turn induce a smearing of the pseudopotential structure factor $S_\kappa(\mathbf{G})$. In a totally similar way as for x-ray diffraction patterns, such a smearing is easily taken into account by introducing a Debye-Waller factor in the structure factor \cite{cardo05a}:
\begin{equation}
S_\kappa(\mathbf{G})\cdot e^{-\frac{1}{2}\left\langle u^2\right\rangle_\kappa\cdot\left|\mathbf{G}\right|^2}\simeq S_\kappa(\mathbf{G})\cdot\left[1-\frac{1}{2}\left\langle u^2\right\rangle_\kappa\cdot\left|\mathbf{G}\right|^2+\cdots\right],
\label{S}
\end{equation}
\noindent where $\left\langle u^2\right\rangle_\kappa$ is the mean-square thermal amplitude of the displacements of the atomic specie $\kappa$. As illustrated in Fig. S\ref{DW-sketch}, the effect of temperature is twofold: i) The static structure factor $S(\mathbf{G})$ is reduced in magnitude by the Debye-Waller factor which contains the mean thermal amplitude of the atomic displacements in the exponent. ii) The lost weight of the peak in $S(\mathbf{G})$ is redistributed forming side tails, corresponding to diffuse scattering processes.

A careful inspection of the empirical pseudopotential results for several $sp^3$ bonded semiconductors \cite{allen81a,gopal87a,cardo89a} led us to the following conclusions regarding the DW term of the electron-phonon interaction: i) For conduction band states which are mostly antibonding in nature, i.e., the wavefunction is mainly distributed around the lattice ions, the DW correction is moderate and negative. This means that the DW term causes a moderate downshift in energy of the conduction band with increasing temperature, corresponding to the reduction of the static structure factor at the reciprocal lattice vectors due to the DW factor. ii) For valence band states which possess mostly bonding character and concentrate the valence charge in the interatomic regions, the DW correction is large and positive. This means that the valence band shifts up in energy with increasing temperature, as a consequence of the increased structure factor weight in the regions between reciprocal lattice vectors due to the appearance of the diffuse-scattering structure-factor tails.

\subsection{The self-energy (SE) correction}

The self-energy correction can be viewed as a change in the effective mass of valence and conduction electrons due to their interaction with the lattice vibrations. The effective mass can be incremented or reduced, for which the electron energy increases or decreases, respectively, depending on the bonding/antibonding character of the states as well as their atomic orbital character. At last, the SE correction depends on how the pseudopotential form factors $V_\kappa(\mathbf{G})$ of each atomic specie react to the different phonon eigenvectors.

From the empirical pseudopotential calculations we infer following trends for the temperature renormalization of gaps in sp$^3$ covalently bonded semiconductors: i) For conduction band electrons the SE correction is always negative, pretty weak for acoustic and strong for optical phonons, and larger for $p$-like states as compared to $s$-like states. This means that conduction band electrons become "heavier" due to their interaction with the lattice vibrations; a coupling that then causes a downshift in energy of the conduction band. ii) For the valence band the sign of the SE correction does depend on the particular phonon modes involved in the interaction. The SE correction is always negative for acoustical phonons but it accidentally almost cancels out with the DW contribution. As a consequence, electron-acoustic-phonon interaction plays a minor role for the gap renormalization in conventional semiconductors. The SE contribution from coupling with optical phonons, in contrast, is negative for $s$-like atomic orbitals but positive (twice as large, though) for $p$-like states. Hence, for conventional semiconductors, for which the top of the valence band and the bottom of the conduction band at the Brillouin zone center has exclusively $p$ and $s$-atomic orbital character, respectively, the fundamental gap decreases with increasing temperature. This is a general trend and as such is considered the {\it normal} temperature dependence of the gap. in frank contrast to that observed for halide perovskites.